# Newly synthesized $Ti_5Al_2C_3$: Electronic and optical properties by first-principles method

M.S. Ali[1], A.K.M.A. Islam[1*], M.A. Hossain[2]

[1]*Physics Department, Rajshahi University, Rajshahi, Bangladesh*
[2]*Physics Department, Mawlana Bhashani Science and Technology, Santosh, Tangail, Bangladesh*

**ABSTRACT**

A theoretical study of the newly identified $Ti_5Al_2C_3$ belonging to the MAX phases has been performed by using the first-principles pseudopotential plane-wave method within the generalized gradient approximation (GGA). The energy band structure and optical properties are reported for the first time. It is seen that Ti $3d$ electrons mainly contribute to the DOS at the Fermi level, and should be involved in the conduction properties. The parameters of optical properties (dielectric function, absorption spectrum, conductivity, energy-loss spectrum and reflectivity) for $Ti_5Al_2C_3$ are calculated and discussed. The material has a large positive static dielectric constant of 130 which indicates it to be a good dielectric material. Further the reflectivity of $Ti_5Al_2C_3$ is high in the infrared-visible-UV region up to ~ 9.7 eV showing promise as a good coating material to avoid solar heating.

## 1. Introduction

Very recently new ternary carbide in the Ti-Al-C system, $Ti_5Al_2C_3$ which belongs to the MAX phases has been identified by Wang *et al.* [1]. The bulk sample was prepared by reactive hot pressing a mixture of Ti, Al and graphite powders with a nominal molar ratio of Ti:Al:C = 5:2.15:2.78 at 1580°C for 1 h. The crystal structure and chemical composition of the new compound have been determined by a combination of XRD and TEM/EDS measurements.

The well-known $M_{n+1}AX_n$ phases (n = 1, 2, 3…) are layered carbides or nitrides with crystal structures of the hexagonal symmetry and have been the focus of large number of investigations (see ref. [2-4]). The MAX phases with larger n value including $(Nb_{0.5}Ti_{0.5})_5AlC_4$ [5] $Ta_6AlC_5$ [6], $Ti_7SnC_6$ [7], and thin film $Ti_7Si_2C5$ [8] were also reported. Besides the above MAX phases, 523 phases have been observed in the deposited Ti–Si–C film [9] and the quarternary V–Cr–Al–C system $(V_{0.5}Cr_{0.5})_5A_2lC_3$ [10]. But the crystal structure of 523 phases has not been reported so far before the work of Wang *et al.* [1].

In this report, we would make a theoretical investigation of the newly synthesized $Ti_5Al_2C_3$ using the first-principles pseudopotential plane-wave method within the generalized gradient approximation (GGA). The electronic band structure and the parameters of optical properties (dielectric function, absorption spectrum, conductivity, energy-loss spectrum and reflectivity) for the phase will be calculated and discussed.

## 2. Computational methods

The zero-temperature energy calculations have been performed using CASTEP code [11] which utilizes the plane-wave pseudopotential based on density functional theory (DFT). The electronic exchange-correlation energy is treated under the generalized gradient approximation (GGA) in the

---

scheme of Perdew-Burke-Ernzerhof (PBE) [12]. The interactions between ion and electron are represented by ultrasoft Vanderbilt-type pseudopotentials for Ti, Al and C atoms [13]. The calculations use a plane-wave cutoff energy 450 eV for all cases. For the sampling of the Brillouin zone, 10×10×2 $k$-point grids generated according to the Monkhorst-Pack scheme [14] are utilized. These parameters are found to be sufficient to lead to convergence of total energy and geometrical configuration. Geometry optimization is achieved using convergence thresholds of $5\times10^{-6}$ eV/atom for the total energy, 0.01 eV/Å for the maximum force, 0.02 GPa for the maximum stress and $5\times10^{-4}$ Å for maximum displacement. Integrations in the reciprocal space were performed by using the tetrahedron method with a $k$-mesh of 54 $k$-points in the irreducible wedge of Brillouin zone (BZ). The total energy is converged to within 0.1mRy/unit cell during the self consistency cycle.

## 3. Results and discussion

### 3.1. Structural properties

The ternary layered carbide $Ti_5Al_2C_3$ with space group $P6_3/mmc$ belongs to the hexagonal systems, the unit cell of which is displayed in Fig. 1. The equilibrium crystal structure of $Ti_5Al_2C_3$ is first obtained by minimizing the total energy. The relevant optimized lattice parameters and internal coordinates are shown in Table 1 along with those obtained by Wang *et al*. [1] in XRD measurement and calculation. The lattice constants and fractional atomic coordinates are seen to be reproduced satisfactorily.

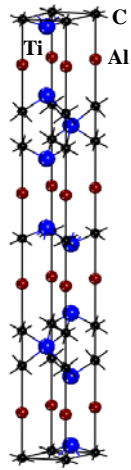

**Fig. 1.** Unit cell of $Ti_5Al_2C_3$.

**Table 1**
The optimized parameters of $Ti_5SiN_4$ compared to XRD data and calculation.

|  |  | Present | XRD [1] | Calc. [1] |
|---|---|---|---|---|
| Lattice parameters | $a$ (Å) | 3.0639 | 3.038 | 3.066 |
|  | $c$ (Å) | 32.7361 | 32.261 | 33.114 |
| Atomic coordinates | $Ti_1$ (4*f*) | (1/3, 2/3, 0.0349) | (1/3, 2/3, 0.0208) | (1/3, 2/3, 0.0340) |
|  | $Ti_2$ (4*f*) | (1/3, 2/3, 0.1778) | (1/3, 2/3, 0.1778) | (1/3, 2/3, 0.1791) |
|  | $Ti_3$ (2*d*) | (2/3, 1/3, 0.25) | (2/3, 1/3, 0.25) | (2/3, 1/3, 0.25) |
|  | Al (4*e*) | (0, 0, 0.1059) | (0, 0, 0.1035) | (0, 0, 0.1059) |
|  | $C_1$ (2*a*) | (0, 0, 0) | (0, 0, 0) | (0, 0, 0) |
|  | $C_2$ (4*e*) | (0, 0, 0.2103) | (0, 0, 0.2102) | (0, 0, 0.2102) |

## 3.2. Electronic properties

Fig. 2 (a) shows the calculated energy band structure of $Ti_5Al_2C_3$ at equilibrium lattice parameters along the high symmetry directions in the Brillouin zone. The valence and conduction bands overlap considerably and there is no band gap at the Fermi level. As a result $Ti_5Al_2C_3$ would exhibit metallic properties.

Fig. 2 (b) shows the total and partial DOS of $Ti_5Al_2C_3$. At the Fermi level $E_F$, the total DOS is 8.2 electrons/eV. From the analysis of the density of states we observe that Ti $3d$ electrons mainly contribute to the DOS at the Fermi level, and should be involved in the conduction properties. The lowest-lying states from about −12 eV to −9 eV are from C $2s$ states. The states from about −8 eV to −5 eV and from −5 eV to −2 eV are mainly from Al $3s$ and C $2p$ states, respectively. A strong hybridization between Ti $3d$ and Al $3p$ states was observed at about −1.8 eV. The hybridization between Ti $3d$ and C $2p$ was observed in the range of −5 to −2 eV. The states in the vicinity of the Fermi level are mainly occupied by Ti $3d$ states. Obviously, Ti $3d$ states give rise to the electrical conductivity in $Ti_5Al_2C_3$. Above the Fermi level, the anti-bonding Ti $3d$ states dominate with less contribution from $3s$ and $3p$ states of Al atoms and $2p$ states of C atoms.

The contribution from Al $3s$ is about an order of magnitude smaller than that of Ti $3d$ states. C $2s$ states, with insignificant contribution from C $2p$ states, do not contribute to the DOS at the Fermi level, and therefore are not involved in the conduction properties. The energy states between -7.5 to 0 eV are due to the hybridizing of Ti $3d$, Al $3s$, Al $3p$, C $2p$ states. The PDOS shows an interesting feature: the hybridization peak of Ti $3d$ and C $2p$ states appear in a lower energy range than that of Ti $3d$ and Al $3p$ states. This suggests that the Ti $3d$-C $2p$ bond is stronger than the Ti $3d$-Al $3p$.

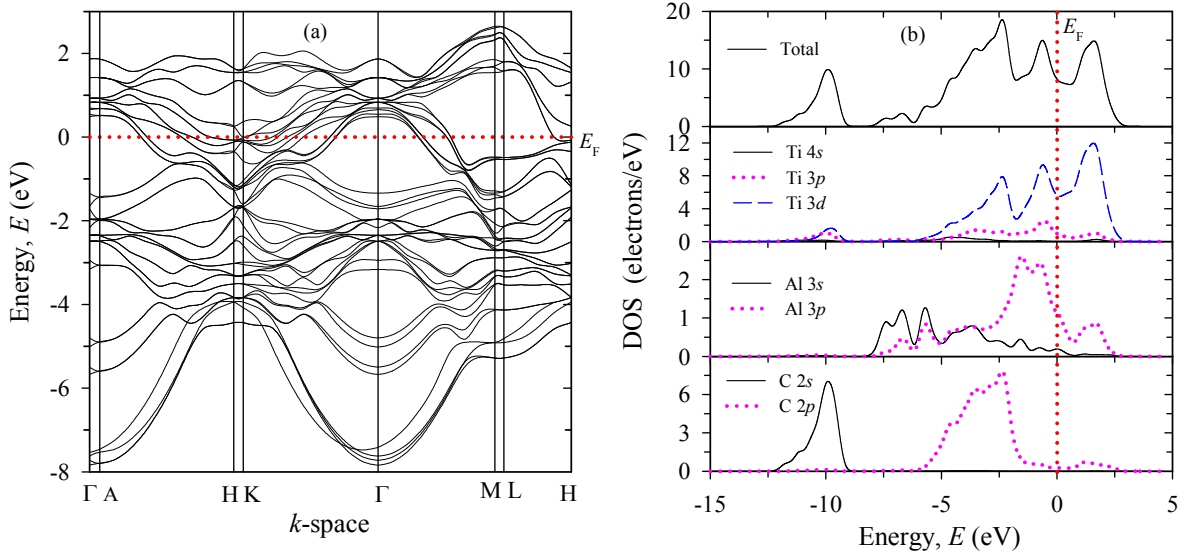

**Fig. 2.** (a) The electronic band structure and (b) DOS of $Ti_5Al_2C_3$.

## 3.3. Optical properties

The study of the optical functions of solids helps to give a better understanding of the electronic structure. The imaginary part of complex dielectric function, $\varepsilon(\omega) = \varepsilon_1(\omega) + i\varepsilon_2(\omega)$, is obtained from the momentum matrix elements between the occupied and the unoccupied electronic states. This is calculated directly using [15]:

$$\varepsilon_2(\omega) = \frac{2e^2\pi}{\Omega\varepsilon_0} \sum_{k,v,c} \left|\psi_k^c \left|u.r\right| \psi_k^v\right|^2 \delta\left(E_k^c - E_k^v - E\right) \tag{1}$$

where $\omega$ is the light frequency, $e$ is the electronic charge and $\psi_k^c$ and $\psi_k^v$ are the conduction and valence band wave functions at $k$, respectively. ***u*** is the vector defining the polarization of the incident electric field. The real part is derived from the imaginary part $\varepsilon_2(\omega)$ by the Kramers-Kronig transform. The expression for all other optical constants, such as refractive index, absorption spectrum, loss-function, reflectivity and conductivity (real part) may be found in ref. [15].

Figs. 3(a)–(f) show the calculated optical properties of $Ti_5Al_2C_3$ from the two polarization vectors (100) and (001) for photon energies up to 20 eV. In our calculation, we used a Gaussian smearing which is 0.5 eV. This smears out the Fermi level, so that k-points will be more effective on the Fermi surface. Despite some variation in heights and positions of peaks, the overall features of the optical spectra for the two polarization directions are roughly similar.

For the imaginary part $\varepsilon_2(\omega)$ of the dielectric function (Fig. 3a), the peak around 0.45 eV is due to transitions within Ti 3*d* bands. On the other hand, $Ti_5Al_2C_3$ has a large positive static dielectric constant $\varepsilon_1(0)$ that is 130, which is close to the value of 126 of $V_4AlC_3$ [16]. In the energy range for which $\varepsilon_1(\omega) < 0$, $Ti_5Al_2C_3$ exhibits the metallic characteristic. The refractive index and extinction coefficient are displayed in Fig. 3 (b). The static value of refractive index is found to be 10.5.

Photoconductivity is the increase in electrical conductivity that results from the increase in the number of free carriers generated when photons are absorbed. We can see from the Fig. 3 (c), the photoconductivity starts at 0 eV which shows that the compound has no band gap. Three peaks are observed at 0.92, 4.3, and 11.56 eV.

The function $L(\omega)$, shown in Fig. 3 (d), describes the energy loss of a fast electron traversing in the material. Its peak is defined as the bulk plasma frequency $\omega_P$, which occurs at $\varepsilon_2 < 1$ and $\varepsilon_1 = 0$ [18, 19]. In the energy-loss spectrum, we see that the plasma frequency $\omega_P$ of $Ti_5Al_2C_3$ is equal to ~9.5 eV. When the incident photon frequency is higher than 9.5 eV, the material becomes transparent.

The absorption spectrum of $Ti_5Al_2C_3$ (Fig. 3 (e)) started at 0 eV due to its metallic nature. It has two peaks, one at 5 eV and the smaller one at 11.3 eV.

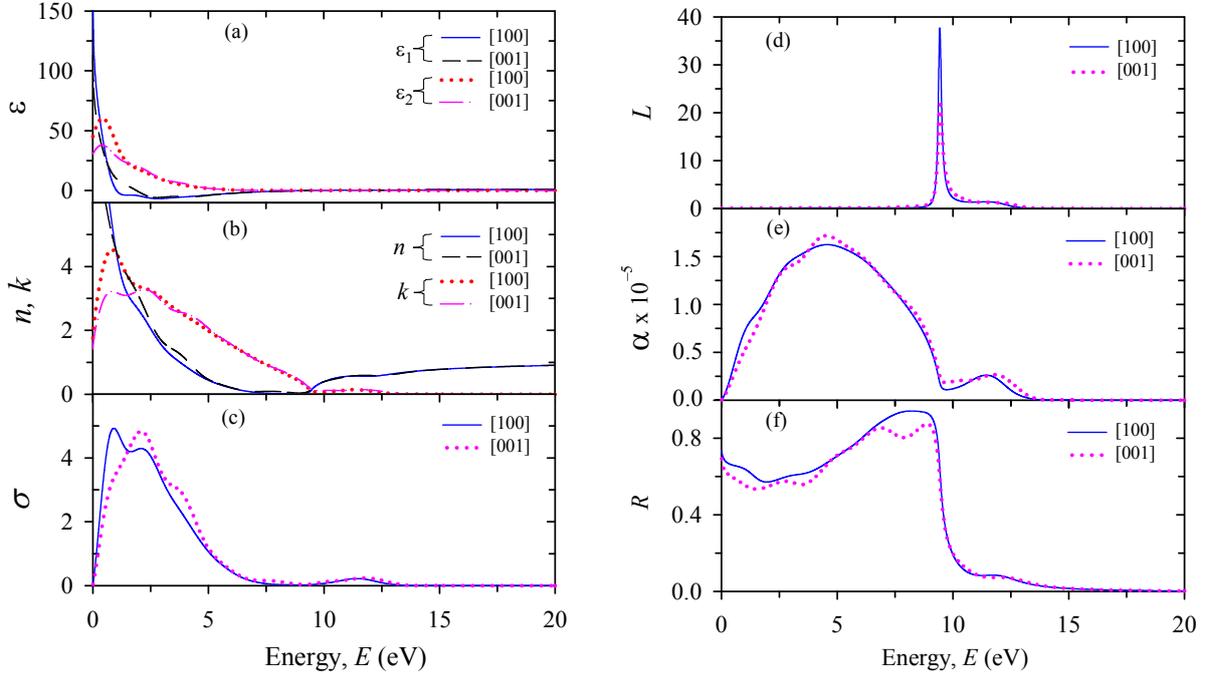

**Fig. 3.** Energy dependent (a) real and imaginary parts of dielectric function, (b) refractive index and extinction function, (c) real part of conductivity, (d) loss function, (e) absorption, and (f) reflectivity of $Ti_5Al_2C_3$ two polarization directions.

The reflectivity as a function of energy is presented in Fig. 3 (f). $Ti_5Al_2C_3$ is seen to possess high reflectivity in the infrared-visible-UV range up to ~9.7 eV. After that the reflectivity falls sharply to low reflectivity (high transparency) for shorter wavelengths. Between 7.5 – 9.4 energy range the ability to reflect solar light reaches maximum of ~ 94%.

## 4. Conclusion

The newly synthesized MAX phase $Ti_5Al_2C_3$ has been theoretically investigated using the first-principles density functional theory within the generalized gradient approximation (GGA). The energy band structure and optical properties are reported for the first time. The conduction properties involve Ti $3d$ electrons which mainly contribute to the DOS at the Fermi level. Further an analysis of optical functions reveals that it a good dielectric material. The reflectivity of the material is quite high in the infrared-visible-UV region up to ~ 9.7 eV showing its capability to reflect solar radiation strongly.